\documentclass{cernyrep} 
\input epsf.tex
\usepackage{epsfig,graphicx}
\usepackage{amsmath,amsfonts,amssymb}
\newcommand{\be}{\begin{eqnarray}}
\newcommand{\ee}{\end{eqnarray}}
\begin{document}

\title{Beyond the Standard Model}

\author{J. D. Lykken}

\institute{Fermi National Accelerator Laboratory, Batavia, IL 60510, USA }

\maketitle 

\begin{abstract}
  Six major frameworks have emerged attempting to describe
particle physics beyond the Standard Model. Despite their different
theoretical genera, these frameworks have a number of common 
phenomenological features and problems. While it will be possible 
(and desirable) to conduct model-independent searches for new physics
at the LHC, it is equally important to develop robust methods to
discriminate between BSM `look-alikes'.
\end{abstract}

\section{Overview of BSM physics} 

``BSM physics'' is a phrase used in several ways. It can refer to physical phenomena 
established experimentally but not accommodated by the Standard Model (SM), in particular dark matter and neutrino
oscillations (technically also anything that has to do with gravity, since gravity is not part of the
Standard Model). ``Beyond the Standard Model'' can also refer to possible deeper explanations of
phenomena that are accommodated by the Standard Model but only with ad hoc parametrizations, such
as Yukawa couplings and the strong $CP$ angle. More generally, BSM can be taken to refer to
{\it any} possible extension of the Standard Model, whether or not the extension solves any particular
set of puzzles left unresolved in the Standard Model.
 
In this general sense one sees reference to the BSM `theory space' of all possible SM extensions, this being
a parameter space of coupling constants for new interactions, new charges or other quantum numbers,
and parameters describing possible new degrees of freedom or new symmetries~\cite{Hubisz:2008gg}.
Despite decades of
model building it seems unlikely that we have mapped out most of, or even the most interesting parts of,
this theory space. Indeed we do not even know what is the dimensionality of this parameter space, or what
fraction of it is already ruled out by experiment. 

Since Nature is only implementing at most one point in this BSM theory space (at least in our
neighbourhood of space and time), it might seem an impossible task to map back from a finite number
of experimental discoveries and measurements to a unique BSM explanation. Fortunately for
theorists the inevitable limitations of experiments themselves, in terms of resolutions, rates, and
energy scales, means that in practice there are only a finite number of BSM model `equivalence
classes competing at any given time to explain any given set of results~\cite{Hubisz:2008gg,LHCinverse}.
BSM phenomenology is
a two-way street: not only do experimental results test or constrain BSM models, they also suggest
--- to those who get close enough to listen --- new directions for BSM model building.

Contrary to popular shorthand jargon, supersymmetry (SUSY) is {\it not} a BSM model: it is a 
symmetry principle characterizing a BSM framework with an infinite number of models. Indeed
we do not even know the full dimensionality of the SUSY parameter space, since this
presumably includes as-yet-unexplored SUSY-breaking mechanisms and combinations of
SUSY with other BSM principles. The SUSY framework plays an important role in BSM physics
partly because it includes examples of models that are `complete' in the same sense as
the Standard Model, i.e., in principle the model predicts consequences for any
observable, from cosmology to $b$ physics to precision electroweak data to LHC
collisions. Complete models, in addition to being more explanatory and making
connections between diverse phenomena, are also much more experimentally constrained than
strawman scenarios that focus more narrowly.

One sometimes hears: ``Anything that is discovered at the LHC will be called supersymmetry.''
There is truth behind this joke in the sense that the SUSY framework incorporates a vast number
of possible signatures accessible to TeV colliders. This is not to say that the SUSY framework
is not testable, but we are warned that one should pay attention to other promising frameworks,
and should be prepared to make experimental distinctions between them. 

Since there is no formal classification of BSM frameworks I have invented my own
(see also Ref.~\cite{Morrissey:2009tf}).
At the highest level there are six parent frameworks:

\begin{itemize}
\item Terascale supersymmetry
\item PNGB Higgs
\item New strong dynamics
\item Warped extra dimensions
\item Flat extra dimensions
\item Hidden valleys
\end{itemize}

Here is the briefest possible survey of each framework, with the basic idea, the 
generic new phenomena, and the energy regime over which the framework purports
to make comprehensive predictions.

\subsection*{Terascale supersymmetry}

The basic idea (for a review, see Ref.~\cite{Chung:2003fi})
is to invoke a new (possibly unique) extension of Poincar\'e spacetime
symmetry to relate elementary excitations of different spins. Thus the generic prediction
is that every SM particle has a superpartner. To explain why superpartners have not
yet been observed, SUSY is spontaneously broken, and it turns out that this
breaking can naturally trigger electroweak symmetry breaking and stabilize its scale.
Supersymmetry models address cosmology because they generically have new very light degrees of freedom
(possibly relevant to inflation and baryogenesis) and with the assumption of conserved $R$-parity
the lightest superpartner (LSP) is stable and could be dark matter. The supersymmetry framework
can also accommodate ultra-high-energy
novelties such as grand unification and strings. Thus the SUSY framework
in principle covers fundamental physics on all energy scales.

Supersymmetry models can be classified according to the dynamics
of supersymmetry breaking, but it is more common to classify them according
to the mechanism by which supersymmetry breaking is {\it mediated} (i.e., coupled) to
the SM sector, shifting the masses of the superpartners of SM particles.
`Direct' mediation implies that the superfields whose vacuum expectation values
break SUSY couple directly to the SM sector, while in `hidden sector' models
the coupling is indirect. Following~\cite{Chung:2003fi} we can divide the hidden
sector models into three groups:  
\begin{itemize}
\item Gravity mediation: Somewhat of a misnomer, since the actual mediation is through
Planck scale suppressed couplings to scalar fields that get vevs due to SUSY breaking
in the hidden sector. One of these scalar fields is the auxiliary field of the $N$$=$$1$ supergravity
multiplet, but in general one needs a theory of physics at the Planck scale to predict all
the couplings. Thus string theory compactifications produce specific gravity-mediated SUSY
models, such as `dilaton-dominated'~\cite{Brignole:1993dj} and $G_2$ models~\cite{Acharya:2008zi}.
Alternatively one can assume by
fiat the absence of many couplings to achieve a simple breaking pattern; examples of this
include `anomaly mediation' ~\cite{Randall:1998uk},  minimal supergravity (mSUGRA)~\cite{msugra},  
gravitino LSP models~\cite{Ellis:2003dn} and `no-scale' models~\cite{Cremmer:1983bf}.
Gravity mediation provides a natural link between
the scale of SM SUSY breaking and the size of the supersymmetric $\mu$ parameter,
but subsumes flavour issues into the mysteries of Planck scale physics.
\item Gauge mediation: SUSY is broken by some new dynamics in a hidden sector,
and communicated to some heavy `messenger' fields that also carry charges under
the SM gauge group (for a review, see Ref.~\cite{Giudice:1998bp}).
SUSY breaking in the SM sector comes from integrating out
these messengers, and is thus suppressed by loop factors as well as by the heavy
messenger mass. Gauge mediation models naturally have a minimal flavour structure
but don't have an obvious explanation for the $\mu$ parameter. Unlike the other 
SUSY breaking masses, the gravitino mass is still Planck suppressed, and thus the
gravitino is the LSP in gauge mediation models; this perhaps makes them less attractive as
explanations of dark matter. 
\item Bulk mediation: Combines SUSY with one of the extra-dimensional frameworks
described below. Some or all of the SM fields are sequestered on a `brane', a kind of slice through the
full `bulk' space that includes the extra dimensions. SUSY is broken by dynamics on
a different brane. The messengers of SUSY breaking to the SM sector are via fields that
propagate in the full bulk space. Examples include gaugino mediation~\cite{gaugoinomed}
and radion mediation~\cite{Chacko:2000fn}.
\end{itemize}

At the end of the day all of these models reproduce the supersymmetrized version of
the Standard Model with some pattern of extra `soft-breaking' masses and
trilinear couplings for the superpartners. If the resulting particle content at the
TeV scale is just this, we have an effective theory of softly-broken SUSY called
the Minimal Supersymmetric Standard Model (MSSM). In particular the Higgs sector
is the minimal one consistent with SUSY: two complex Higgs doublets leading to
three neutral Higgs bosons and two charged ones. I know of no reason to assume
that this SUSY Higgs sector should be minimal. This nomenclature is also agnostic
about the gravitino, which certainly matters if it is the LSP. Often one sees references
to the `constrained' MSSM (CMSSM)~\cite{Kane:1993td},
which is just the low-energy SM sector arising
from the mSUGRA scenario. From a phenomenological point of view it is more
generic to use a less-constrained parametrization, such as the `pMSSM'~\cite{Berger:2008cq}.

\subsection*{PNGB Higgs}

The basic idea is that the Higgs boson is a pseudo-Nambu--Goldstone boson of
some global symmetry that is both spontaneously broken by some new physics
at $\sim 10$ TeV and explicitly broken because part of the global symmetry is gauged to become
the weak interactions. Thus in this framework the Higgs is naturally light compared
to the 10 TeV scale for essentially the same reason that the pion in QCD is light
compared to the masses of the proton and $\rho$ meson. 
For reviews see Ref.~\cite{LH}.

The explicit breaking of the global symmetry requires extra assumptions to avoid
quadratic divergences from loops of $W$'s, $Z$'s, and tops that would destabilize the weak scale. 
The generic solution
is to introduce heavy partners for SM particles and arrange for cancellations of
divergences at one loop. In one successful variant, Little Higgs with $T$-parity,
every SM particle has a heavier partner with the same spin and charges, while
the top quark has extra partners; the lightest partner particle is stable due to
$T$ parity and makes a good dark matter candidate. In another variant, Twin Higgs,
the partners have no SM charges but form a mirror copy of the SM under some
hidden gauge interactions~\cite{Chacko:2005pe}.

The PNGB Higgs framework is designed to explain particle physics 
comprehensively up to a cutoff energy scale of about 10 TeV.
At the cutoff one should replace each PNGB model by some
`UV completion', i.e., a high-energy model possibly drawn from
one of the other frameworks discussed here: supersymmetry, new strong
interactions, or warped extra dimensions.

\subsection*{New strong dynamics}

In the Landau--Ginzburg theory of superconductivity, the superconducting state is
characterized by a charged bosonic order parameter;
in BCS theory this picture is only an approximation to a
microscopic dynamical condensate of electron pairs. It is possible that the same holds
for Higgs theory, with electroweak symmetry broken by fermion condensates
bound by some new strong interaction. The simplest version of this idea, called technicolour,
assumes a new force similar to QCD but with a much larger characteristic energy scale,
and technifermions that carry both electroweak charges and technicharges.
In other versions, called top condensation or topcolour, the electroweak breaking
condensate involves top quarks. For a review, see Ref.~\cite{Hill:2002ap}.

Because strong dynamics is involved, it is difficult to determine directly if
there are actually any realistic models in this framework. Furthermore, once
the Higgs has been replaced by a condensate, one faces the additional challenge
of generating realistic SM quark and lepton masses without benefit of SM
Yukawa couplings, i.e., one is forced to construct a realistic dynamical theory
of flavour. Models of new strong dynamics typically do not consider physics
at scales above 10 to 1000 TeV, unless they are combined with supersymmetry.

\subsection*{Warped extra dimensions}

The basic idea is that there is an extra spatial dimension with a strongly
warped geometry, such that high energy scales as measured on one
slice of the fifth dimension rescale to low energy scales as measured on
another slice. In particular the Planck scale (or superstring scale) could
be $10^{19}$ GeV on one slice, but only a few TeV on 
another~\cite{warped}.

Because of the warped geometry each SM particle has to be described
by a wave function describing how it is spread out or localized in the
fifth dimension. Localizing the Higgs where the ultimate UV cutoff
is only a few TeV helps explain the weak scale -- Planck scale hierarchy,
while localizing the SM fermions to have different overlaps in the
fifth dimension gives a natural theory of flavour.

SM particles and/or the graviton have heavy Kaluza--Klein (KK) partners that
could be produced at the LHC. Realistic warped models typically introduce
additional new particles. The Higgs sector is potentially richer, either because
of mixing with the scalar `radion' whose mass stabilizes the warped geometry,
or because the Higgs is secretly the fifth component of a five-dimensional gauge
field.

Because of AdS-CFT duality in string theory, it is possible that some/most/every
warped model can be promoted to a more complete model that in turn has a dual
low energy description as four-dimensional model of new strong dynamics.
There are also Higgsless warped models~\cite{Csaki:2005vy},
in which the role of the Higgs
is replaced by a combination of new strong dynamics with perturbative
effects of the KK partners of the $W$ and $Z$.

\subsection*{Flat extra dimensions}

The basic idea is that there are flat extra spatial dimensions of finite extent
with special four-dimensional slices called branes. If we imagine that all
SM particles are confined to a brane, then we are in the ADD scenario~\cite{ArkaniHamed:1998rs}
and the extra dimensions could have macroscopic size (a few microns).
If we imagine that the SM fields extend through all the dimensions
(with perhaps some special interactions limited to branes), then we
are in the scenario of Universal Extra Dimensions~\cite{UED}.

In realistic 5d and 6d
models of UED, all of the SM particles have heavy KK partners.
The lightest KK partner is stable, due to a remnant of momentum
conservation in an extra dimension, leading to a good dark matter
candidate. Note this `KK-parity' is not ad hoc,
unlike its counterparts $R$-parity and $T$-parity.

UED and ADD models (and combinations thereof) do not attempt
to explain the dynamical origin of their own geometries and
energy/length scales. Some models of TeV-scale extra dimensions
provide novel mechanisms for electroweak symmetry breaking.
The most extreme version of ADD is a
strawman scenario for black hole production at the LHC.

\subsection*{Hidden valleys}

The other frameworks described above sometimes invoke hidden sectors with
some essential new dynamics (SUSY breaking, mirror gauge groups, etc.).
One can just as well use the assumption of one or more hidden sectors
as a framework in its own right~\cite{Strassler:2006im}.
The basic idea is that there are one or many
new particles that are relatively light --- perhaps as light as or lighter than SM particles --- but
difficult to detect because they interact only weakly with SM particles and may
be subject to new conservation laws. This hidden sector would generically
have its own gauge interactions; these could be mixed slightly with SM
interactions due for example to quantum mixing effects from very heavy particles that carry
both kinds of charges. The hidden sector gauge theory might also be exotic,
e.g., it could respect conformal invariance up to a small breaking due to
its coupling to the SM~\cite{Georgi:2007ek}.

In recent years the most interesting developments in this framework are related to
dark matter. Here the relevant starting point is to ask what are the particles/interactions
that {\it mediate} between dark matter particles (or their relatives
in the dark sector) and SM particles~\cite{Bai:2009ms}.
The answer could be only gravitational
or other ultra-suppressed interactions, and the answer could be the
weak mediators of the SM ($Z$'s and/or Higgs). The third option,
currently popular in attempts to reconcile various possible signals
in direct or indirect dark matter detection, is exotic mediators such as
a new light gauge boson with a mixing-suppressed coupling to 
SM matter~\cite{ArkaniHamed:2008qn}.

The hidden valley framework is the most `bottom-up' of the BSM frameworks,
meaning that model-building it is driven by phenomenological considerations
rather than `top-down' invocations of new principles of fundamental physics.

\section{Common problems and features of BSM models}

Even from this minimal pr\'ecis the alert reader may have detected some phenomenological
similarities in BSM models arising from very different theoretical frameworks.
These similarities may to some extent derive from deep connections between the
frameworks, such as AdS/CFT duality, but from a more pedestrian model-builders'
perspective they arise from the necessity of evading strong constraints on BSM
models coming from existing data. 

\begin{figure}[t]
\center
\psfig{file=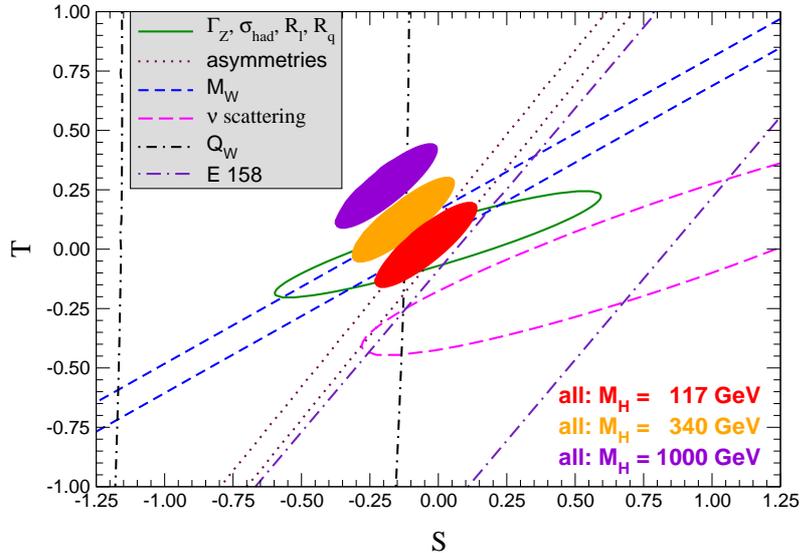,width=9cm,angle=-90}
\caption{The shaded ovals show the 90\% CL experimentally allowed regions for
the $S$ and $T$ parameters, for three different assumed values of the
Higgs boson mass. The dotted and dashed lines are constraints from
various particular measurements. Taken from Ref.~\cite{pdg}.
}
\label{fig:STplot}
\end{figure}

Precision measurements from experiments at LEP, the Tevatron, the SLC, the $B$ factories,
and with kaons, as well as searches for rare processes such as proton decay or charged
lepton number violation, or for $CP$-violating electric dipole moments, put strong limits on
new physics phenomena at multi-TeV energy scales and higher. BSM models from
many well-motivated frameworks often get into trouble with these limits, usually for
one of two reasons:
\begin{itemize}
\item The Standard Model has three accidental global symmetries: $B-L$ is exactly
conserved, $B+L$ is conserved up to nonperturbative effects, and custodial symmetry
--- the diagonal remnant of $SU(2)_L\times SU(2)_R$ ---  is conserved up to
effects from the SM up-versus-down Yukawa splittings. In addition, the SM has no
flavour-changing neutral currents at tree level, and a very strong GIM suppression
of FCNCs at loop level. BSM models generically violate one or more of these
symmetries, sometimes at tree level (operators of dimension four) and sometimes
by generating higher dimension operators with insufficient suppression.
\item In attempting to make BSM models that explain/stabilize the weak scale,
one almost inevitably introduces new particles that carry electroweak charges,
or extend the electroweak sector. Generically this creates conflict with, for example, the
Peskin--Takeuchi precison observable $S$, which roughly speaking counts the
number of left- versus right-handed fermion weak doublets, the observable $T$, which is 
sensitive to large mass splittings between weak isospin partners, and the
$Zb\bar{b}$ vertex, which is sensitive to new heavy fermions that can contribute
to the vertex at one loop. The current $S$--$T$ constraints are shown in Fig.~\ref{fig:STplot}.
\end{itemize}
In turn there seem to be three general strategies to avoid such problems in
BSM model building:
\begin{itemize}
\item Make the new particles heavy, where heavy means 2 or 3 TeV in favourable
cases, and tens of TeV or more in unfavourable cases. 
\item Introduce new symmetries to forbid tree-level contributions of
new particles to SM processes, for example. Thus  $R$-parity in SUSY,
$T$-parity in Little Higgs, and KK-parity in UED force new particles to
be created in pairs.
\item Suppress the couplings between the new particles and SM particles.
This of course leads to the hidden valley framework. Note also that the
most straightforward mechanisms to motivate such small couplings are 
to make at least some exotic particles very heavy, or to invoke new
symmetries; so this third approach overlaps with the first two.
\end{itemize}

The net result of all this is that successful BSM models tend to be either
`heavy', `hidden', or efficiently decoupling. The `heavy' models are somewhat
discouraging in that they put much of the new physics out of reach of even the LHC,
and they suffer also from the {\it little hierarchy problem:} if the new physics is
supposed to have the virtue of stabilizing the weak scale, how do the new particles
associated with this physics manage to have masses much larger than $M_Z$
and the Higgs mass? The `hidden' models may also put much of the new physics 
out of reach of the LHC, but this depends on details of the model and the `hiding';
still this raises the piquant question of whether the physics behind dark matter
will turn out to be accessible to colliders. 

SUSY is the prime example of an
efficiently decoupling model, and in fact SUSY models with relatively light
superpartners fit the precision data slightly {\it better} than the Standard Model.
However, SUSY models also suffer at least somewhat from the little hierarchy
problem, since they relate $M_Z^2$ to a linear combination of superpartner
masses-squared with (model-dependent) coefficients of order one; thus the
absence of a $\sim 100$ GeV chargino at LEP and a $\sim 200$ GeV gluino
at the Tevatron creates some tension. In the MSSM this is much worse, because
the assumption of a minimal Higgs sector would have made the Higgs boson
accessible at LEP, absent large radiative corrections to its mass from heavy
($\sim 1$~TeV) superpartners.

\section{How to make new particles at the LHC}

\begin{figure}[htbp]
\begin{center}
\includegraphics[width=0.290\textwidth]{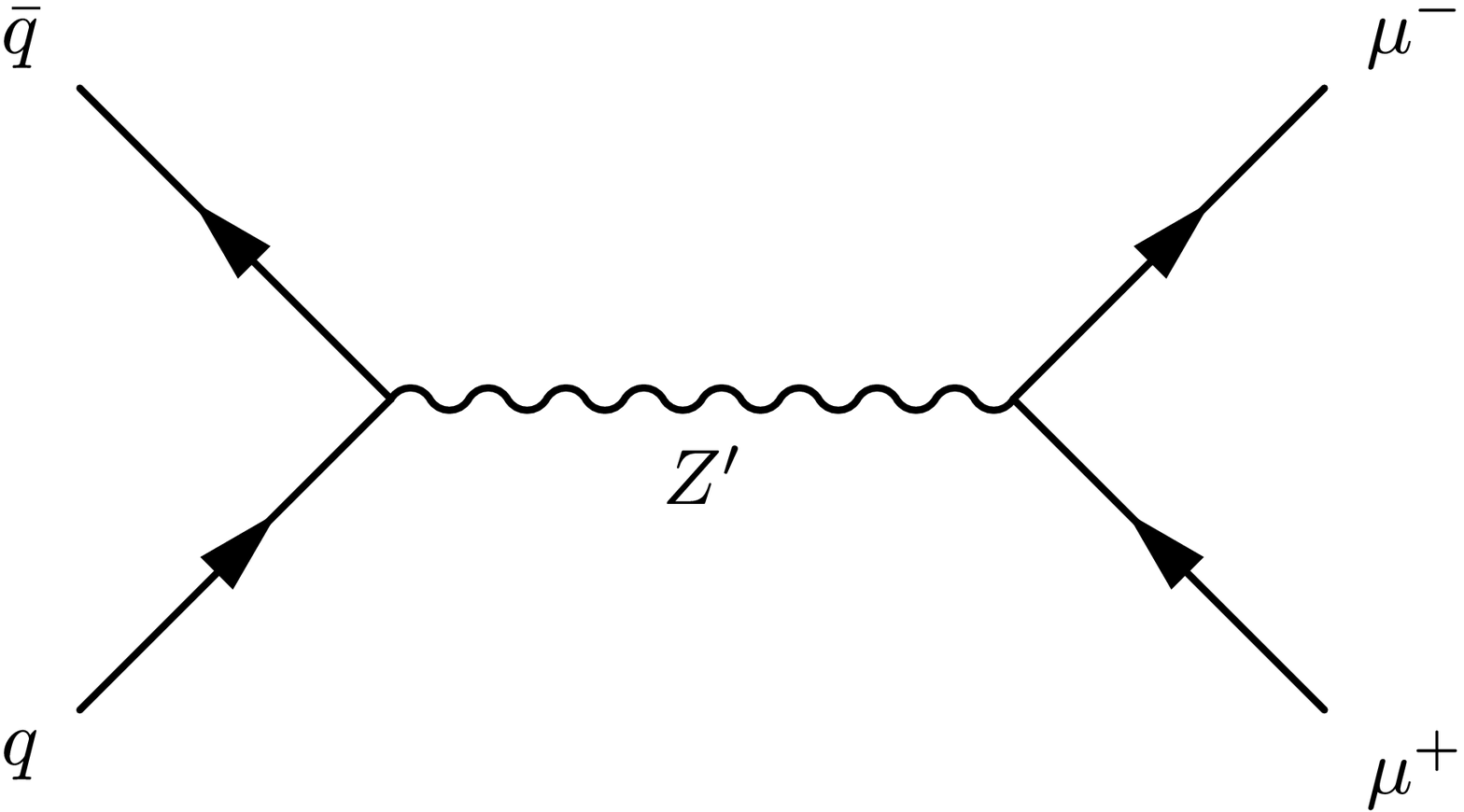}\hspace*{20pt}
\includegraphics[width=0.290\textwidth]{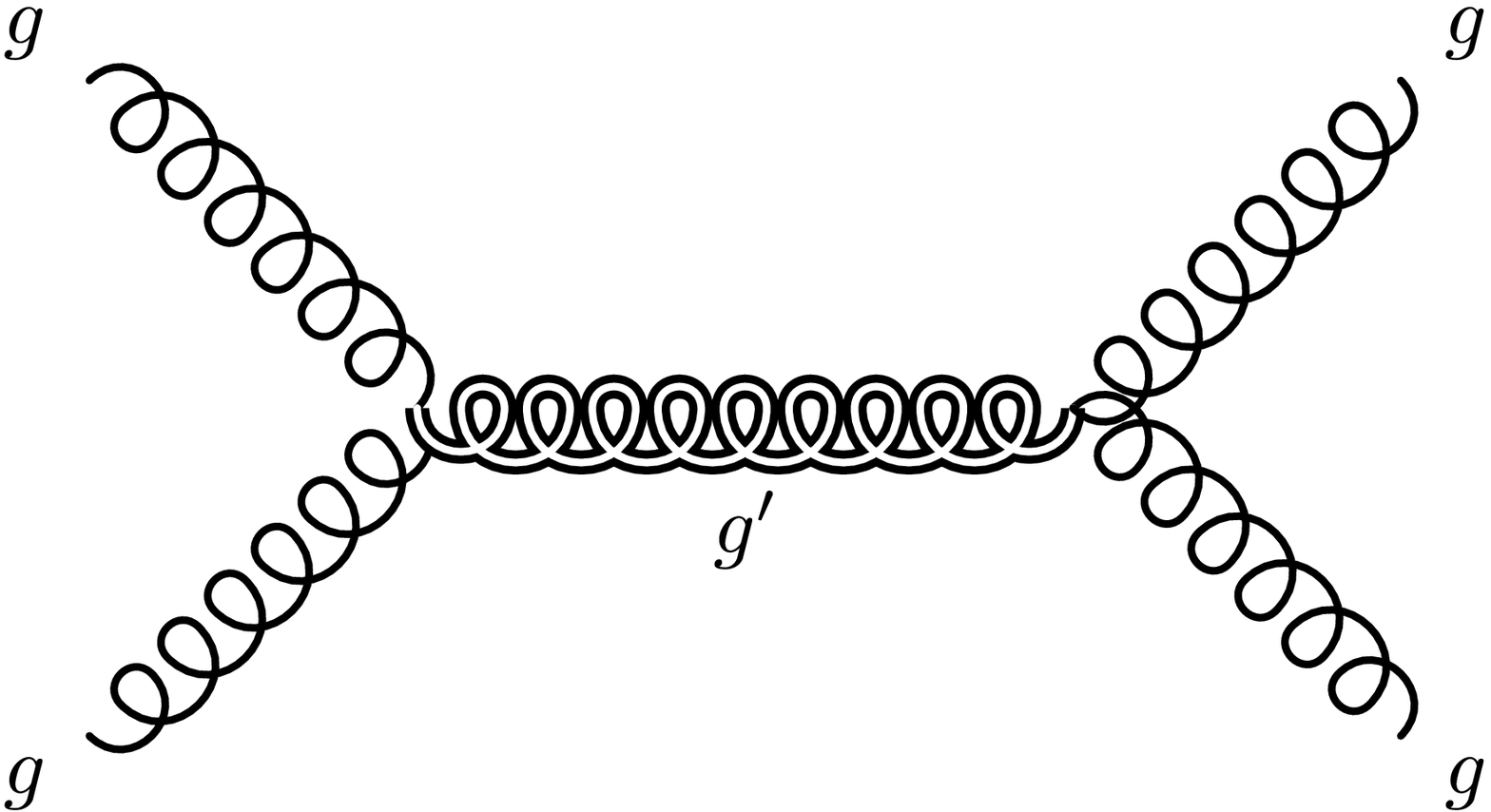}\hspace*{20pt}
\includegraphics[width=0.290\textwidth]{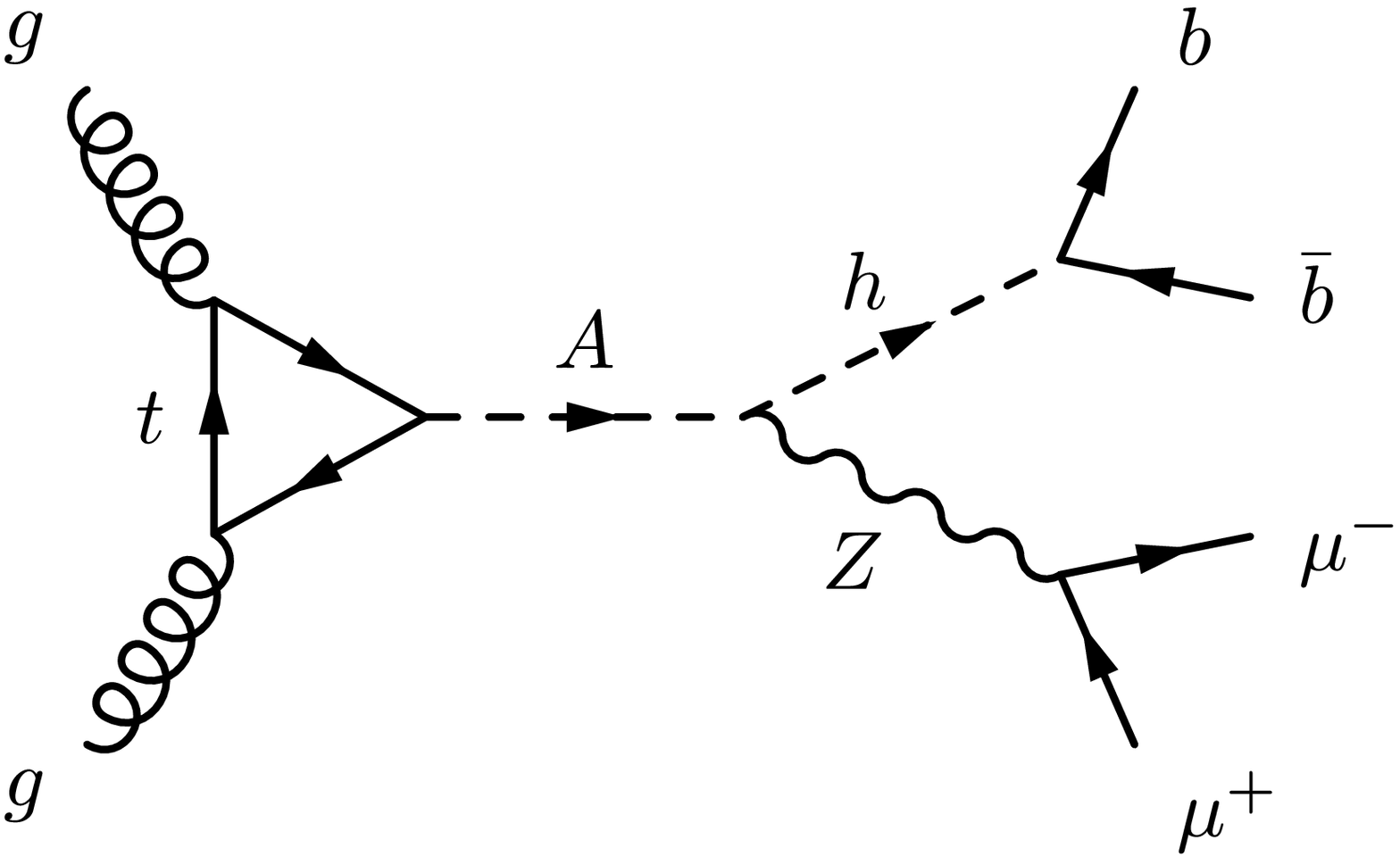}
\caption{Diagrams for resonant production of an exotic particle:
a $Z'$ gauge boson (left), a Kaluza--Klein massive gluon (centre), and a heavy
pseudoscalar (right) \label{fig:schanprod}}
\end{center}
\medskip
\begin{center}
\includegraphics[width=0.290\textwidth]{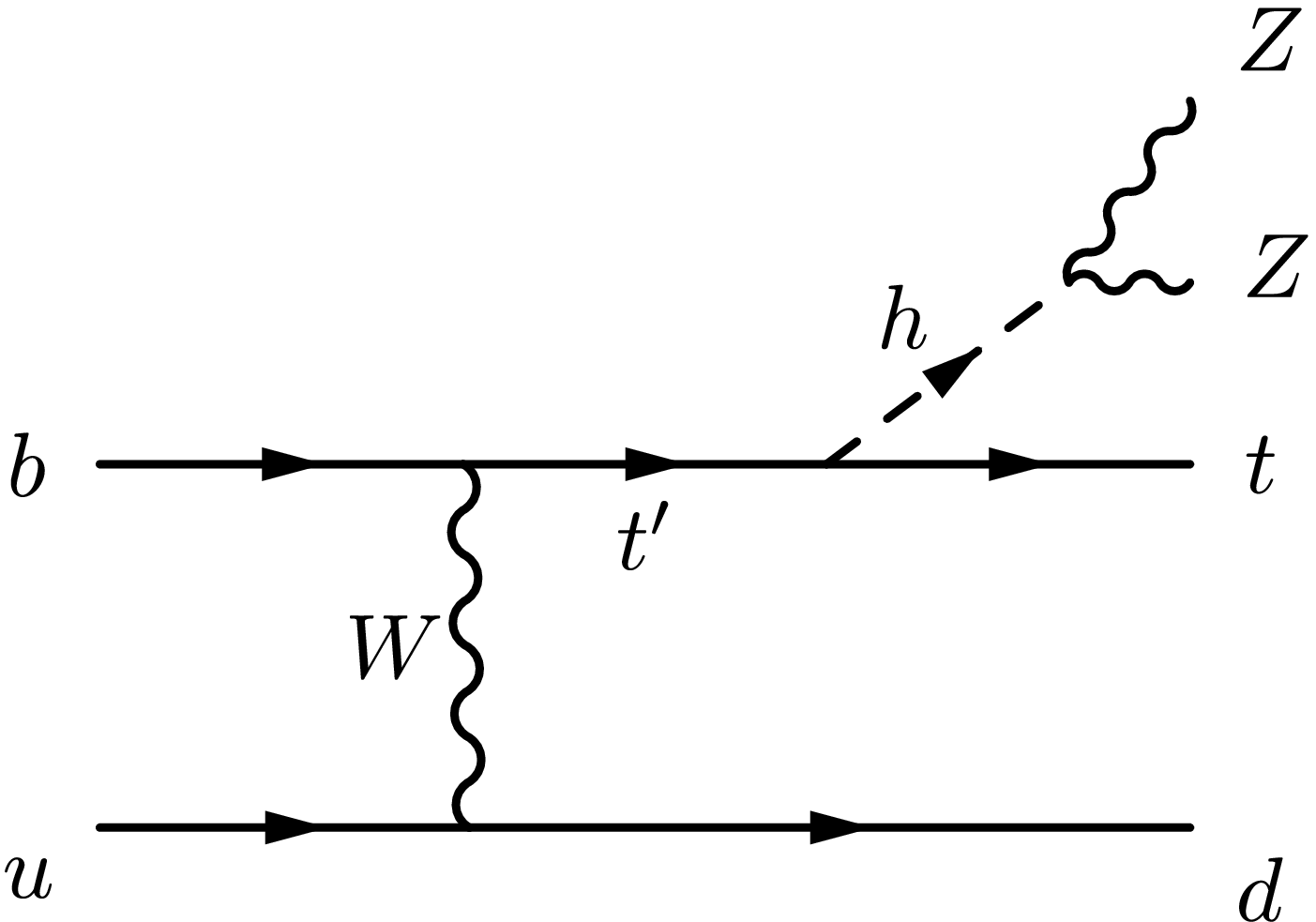}\hspace*{20pt}
\includegraphics[width=0.290\textwidth]{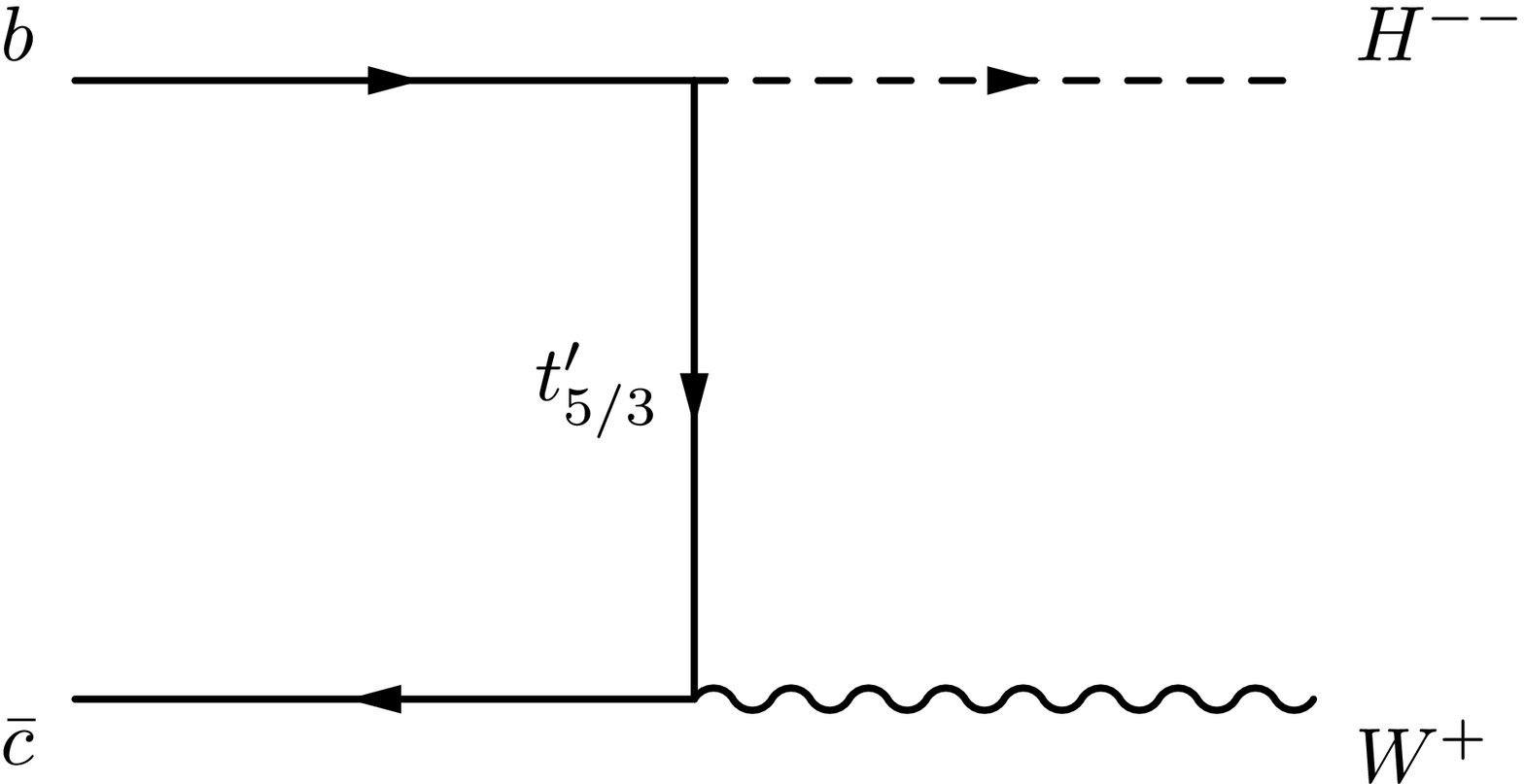}\hspace*{20pt}
\includegraphics[width=0.290\textwidth]{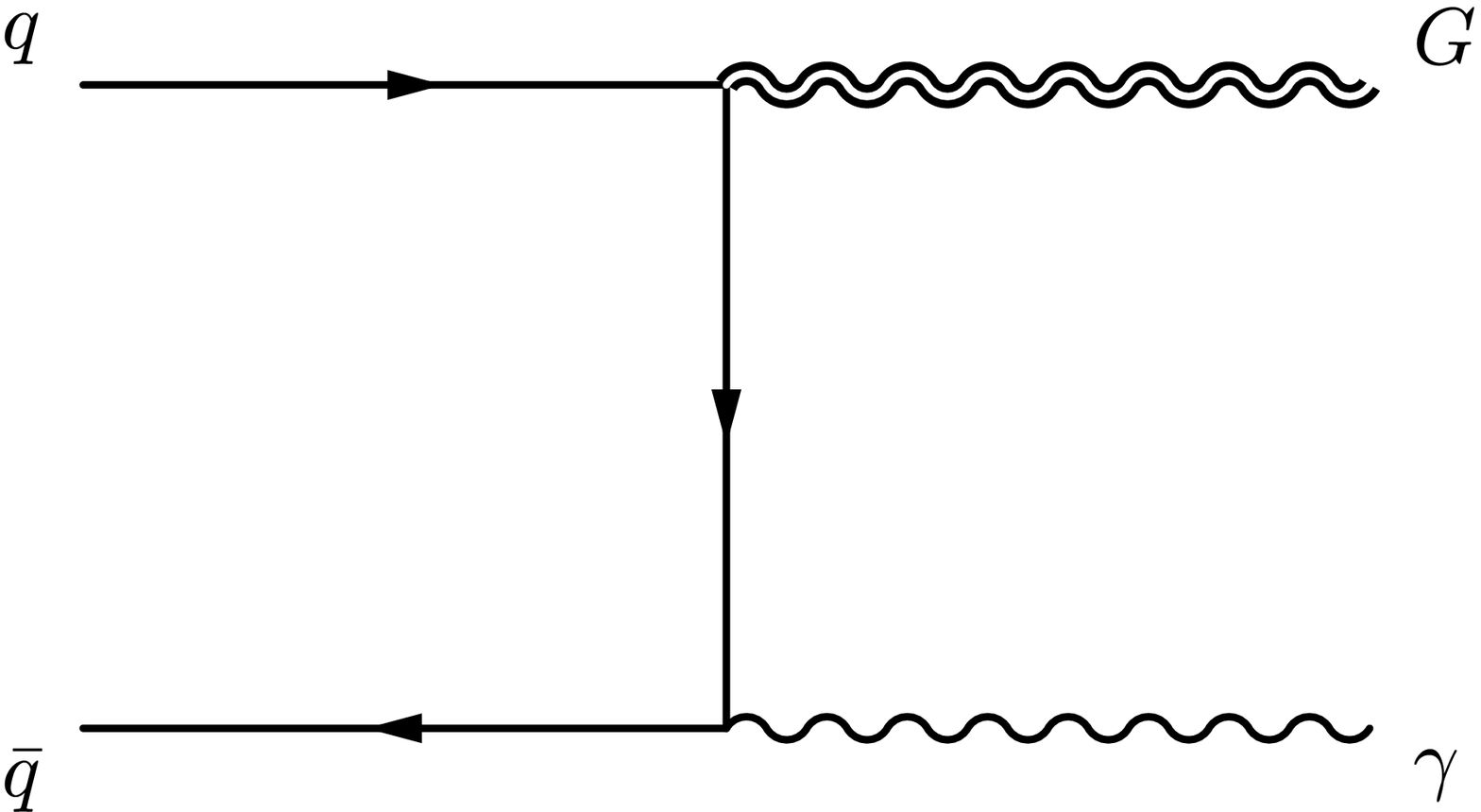}
\caption{Diagrams for associated production of an exotic particle:
a $t'$ heavy quark (left), a doubly charged scalar (centre), and a Kaluza--Klein graviton (right) \label{fig:assocprod}}
\end{center}
\end{figure}

\begin{figure}[htbp]
\begin{center}
\includegraphics[width=0.450\textwidth]{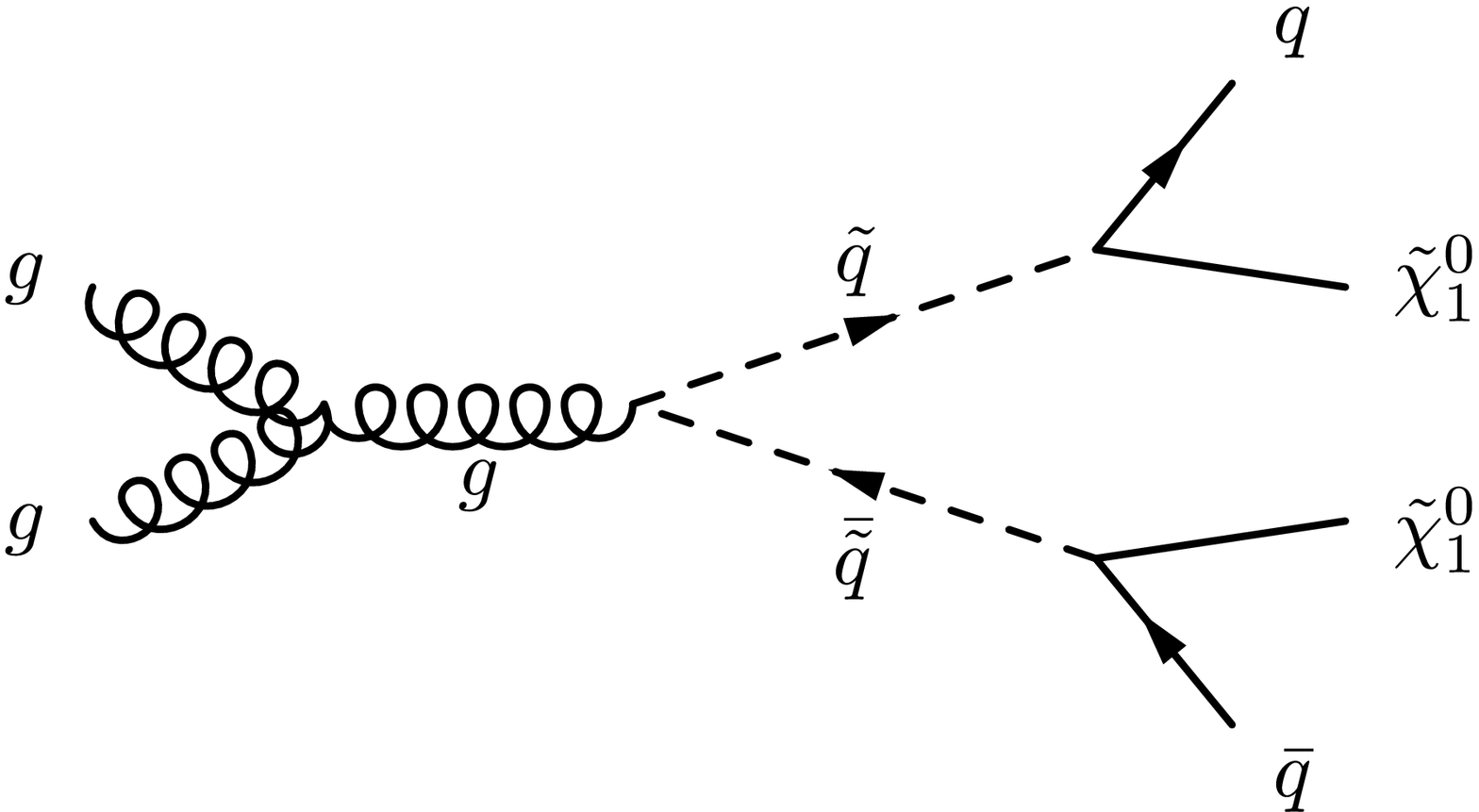}\hspace*{20pt}
\includegraphics[width=0.450\textwidth]{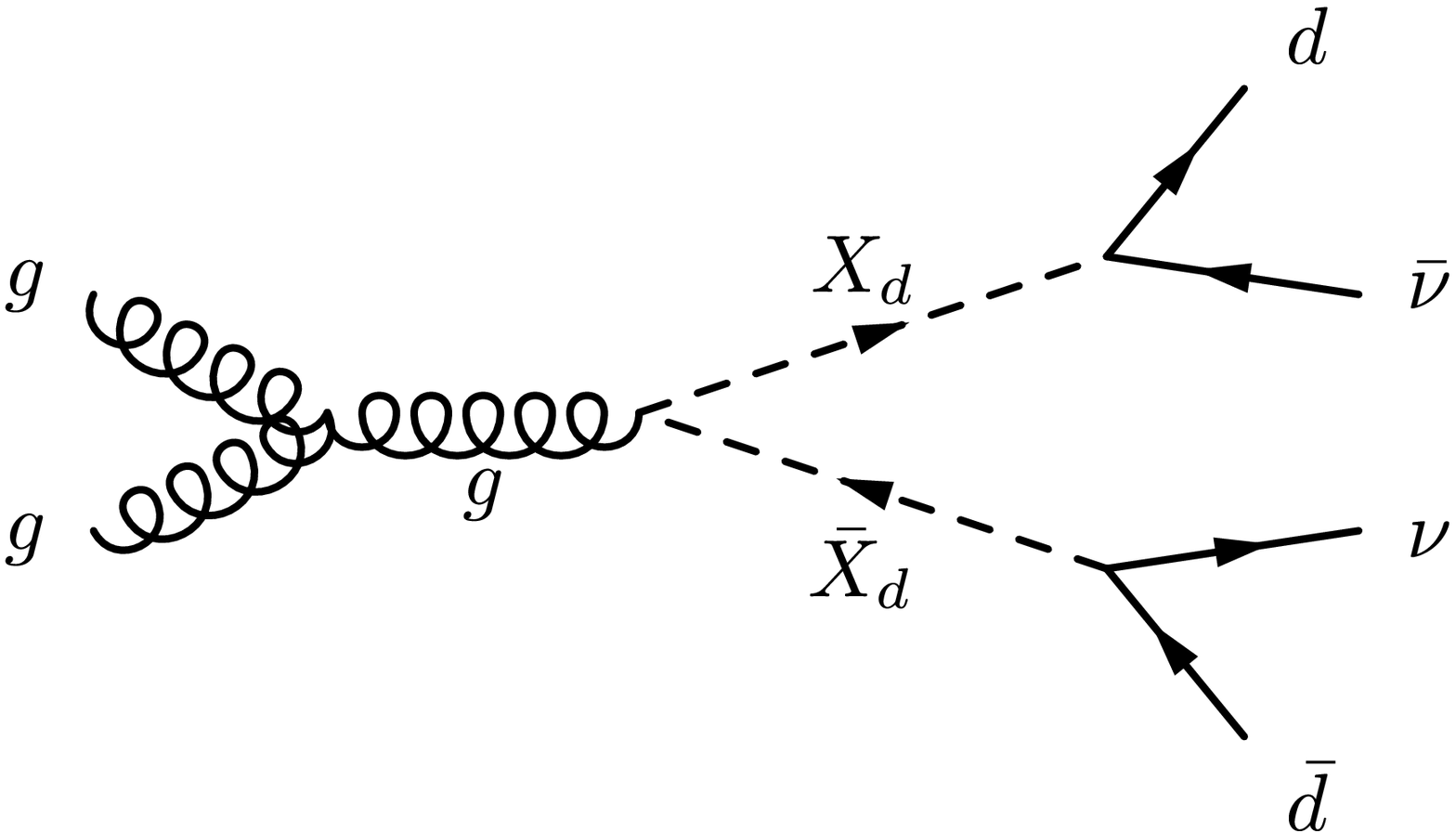}
\caption{Diagrams for pair production of exotic particles: squarks (left) and leptoquarks (right) \label{fig:pairprod}}
\end{center}

\begin{center}
\includegraphics[width=0.480\textwidth]{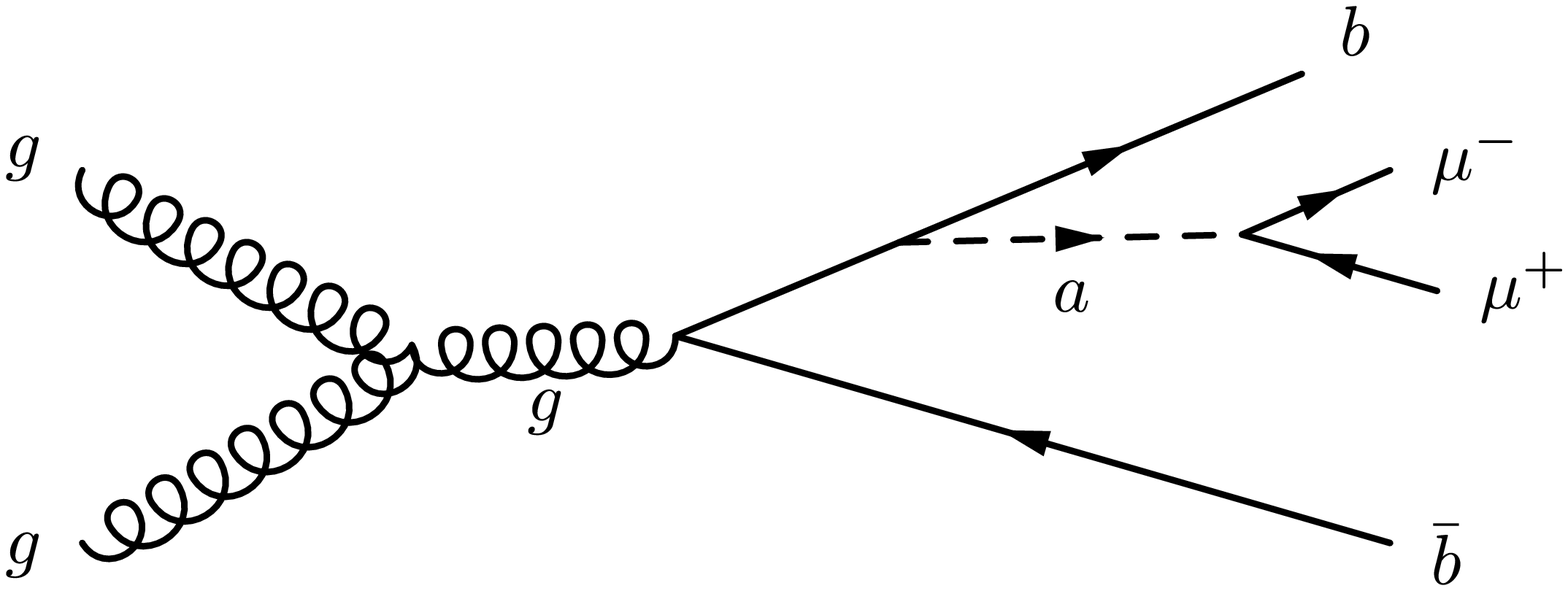}\hspace*{20pt}
\includegraphics[width=0.400\textwidth]{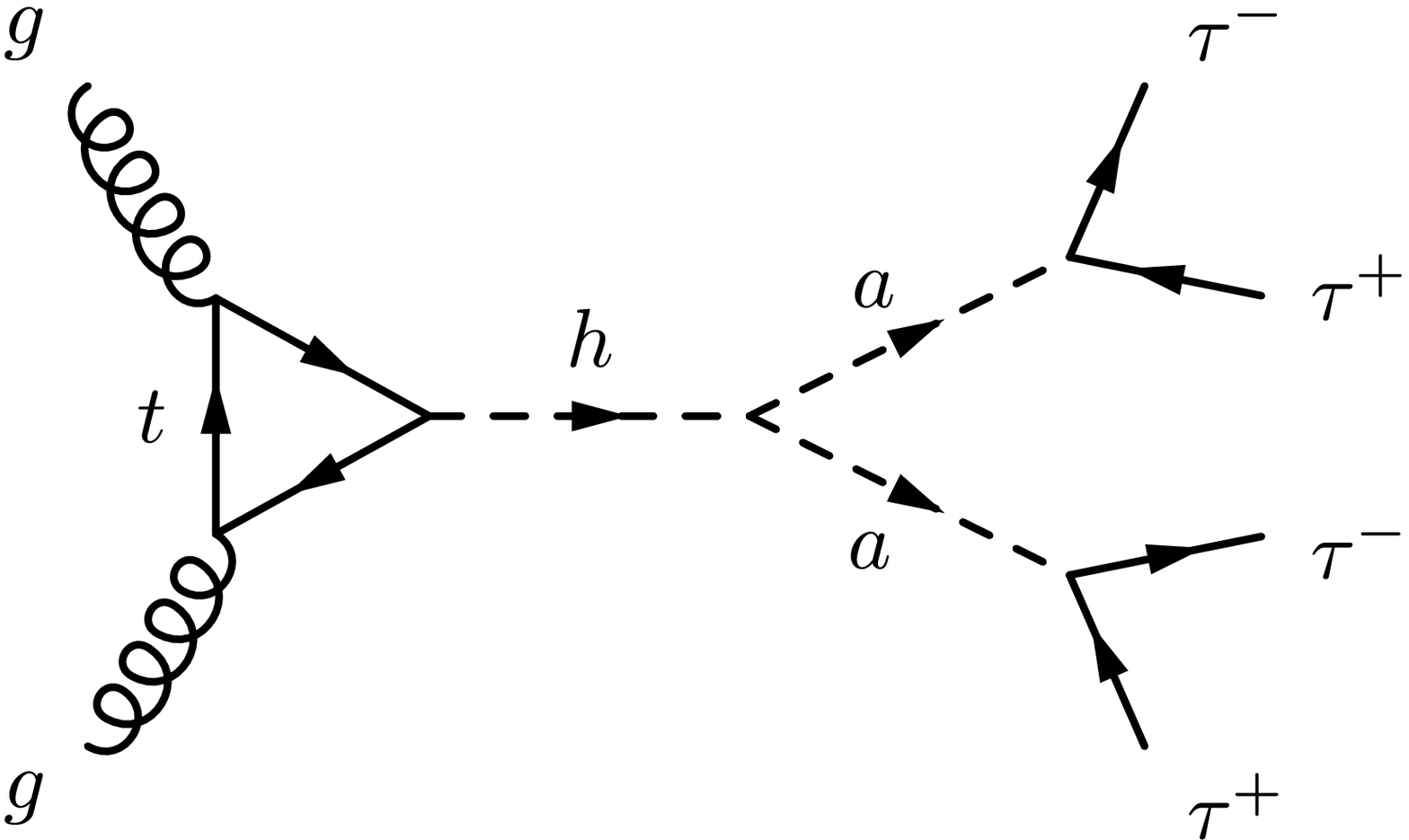}
\caption{Diagrams for BSMsstrahlung production of an exotic pseudoscalar particle (left) and
exotic peusdoscalars from the decay of a Higgs boson (right) \label{fig:decayprod}}
\end{center}
\end{figure}

Since the LHC is an energy-frontier collider, we hope that  LHC discoveries 
at ATLAS and CMS will
arise mostly from the production and decay of some new on-shell resonances
(as opposed to virtual effects on SM processes).
Putting aside the more exotic possibilities of producing extended objects
(e.g., superstrings, quirks) or black holes, the ATLAS/CMS discovery programme
is to look for new particles. Here is a more-or-less comprehensive
model-independent list of all possible discovery modes:
\begin{itemize}
\item A new $s$-channel resonance, with decay to a pair of SM particles,
a pair of lighter exotics, to one of each, or more complicated cascades.
\item Single production of a new heavy particle in association with an
SM particle.
\item Pair production of exotics, with decay chains that end in either all
SM particles or include one or more stable or quasi-stable exotics.
\item BSMstrahlung, i.e., a high-energy SM process with extra radiation of
one or more exotic particles.
\item Exotics produced in the decays of `Standard Model' particles, probably
top or Higgs.
\end{itemize}
A few examples are illustrated in Figs.~\ref{fig:schanprod}--\ref{fig:decayprod}.

An LHC discovery search plan at some level
is just making a list of possible final states, performing a large number of
careful inclusive searches in different channels, and studying unexplained excesses.
Thus to first approximation LHC experimenters do not need to know anything about
BSM models in order to make discoveries  --- but they need to know a lot about Standard Model physics!

\section{Look-alikes at the LHC}

Suppose you are fortunate enough to see a new resonance in one or more
channels at ATLAS or CMS. Now you would like to determine what kind of
a beast have you found, and what the existence of this exotic tells us about
physics beyond the Standard Model.

This task, I would claim, is best considered as a problem of discriminating between 
{\it look-alikes}. At the purely phenomenological level the look-alikes can be 
particles that differ in their basic quantum numbers but produce the same
final state. A good example of this type of look-alike problem is 
Higgs discovery~\cite{DeRujula:2010ys,Gao:2010qx}:
just because you found a resonance that decays into, e.g., a pair of $Z$ bosons,
doesn't mean that it is the neutral spin 0 $CP$-even member of a ($2_L$, $2_R$)
doublet under $SU(2)_L\times SU(2)_R$. Your new resonance could instead have
spin 2, or be a pseudoscalar, or belong to a ($4_L$, $4_R$) of $SU(2)_L\times SU(2)_R$.
Before booking a trip to Stockholm you might want to see if your data can
discriminate between these hypotheses and the Higgs boson of the Standard Model.

An even greater challenge occurs if you make a `missing energy' discovery, i.e., you
establish an excess of unbalanced high-energy events with large missing transverse
energy. Any such discovery could be related to dark matter, but a long chain of reasoning
and additional measurements --- not just at colliders --- will be required to make this
connection. To begin with, one would like to determine how many invisible particles
are produced per event (1? 2? 26?) and whether these unseen particles are heavy
or light or are just neutrinos. If the invisible particles are produced from decays of other exotics
or in association with other exotics, we would like to determine the quantum numbers
of these particles as well. With some of this basic information in hand, we then want
to map the missing energy signature back to one of the BSM frameworks. This
is challenging because, as we saw, several frameworks produce similar signatures
of pair-produced new heavy particles cascading to pairs of weakly-interacting stable 
(or quasi-stable) exotics~\cite{Hubisz:2008gg},~\cite{Cheng:2002ab}--\!\cite{Hallenbeck:2008hf}.

A major advantage of thinking in terms of look-alikes is that, at any given time,
you reduce your experimental problem to hypothesis testing: Does the data
prefer A or B? More precisely, you can answer a question like ``Given that Nature
chose hypothesis A, how likely is it that my experimental data would nevertheless
favour hypothesis B?'' This strategy is very powerful because it allows using all of the
information from your data, and because the number of successful hypothesis tests needed 
to discriminate among $N$ possibilities in BSM theory space scales like log$N$.
The difficulty of this approach is that you need to know a lot about BSM physics,
enough to both navigate the BSM theory space effectively and to compute likelihoods
for LHC datasets given a particular BSM origin.

\section{What is missing?}

All existing BSM models that strive for some kind of comprehensive extension
of the SM are unsatisfactory in some way. Most resort to some otherwise unmotivated
tunings or adjustments of mass scales to avoid conflicts with data, and others are in
significant tension with data. As BSM model-builders try to explain more or to be more
explicit about the underlying structure and details of their models, the models themselves tend to 
accumulate more jury-rigged features, ending up somewhere between Baroque
and Rococo in appearance. This does not imply that the basic BSM ideas are
incorrect, but it may imply that we are all missing some essential ingredients in
our model-building.

What is missing? Perhaps our implementation of supersymmetry or the
basic idea of `partners' is too naive. Perhaps we need to `deconstruct'~\cite{ArkaniHamed:2001nc}
our thinking about degrees of freedom in general, not just for extra dimensions.
Perhaps everything comes ultimately from strong dynamics and we just
don't have the tools to figure out the rules. Perhaps gauge theories and/or
quantum field theory is just an emergent approximation for a deeper
underlying framework. Perhaps the Standard Model 
(tweaked to accommodate neutrino masses) is all there is, the parameters
of the SM are determined by initial conditions of the Big Bang, dark matter
is all black holes, and the accelerating expansion of the observable
universe has a purely macroscopic origin.

What is missing? Enough data from enough experiments to point us
in a clear direction, and start to resolve at least some of our long-standing
puzzles. We hope that the advent of the LHC, along with exciting results
from astrophysics, direct dark matter detection, neutrinos, and searches for rare
processes, will do the job.


\section*{Acknowledgements}
 Thanks to Gustavo Burdman, Marcela Carena, 
Alvaro De~R\'ujula, Keith Ellis, Patrick Fox, Maurizio Pierini, Maria Spiropulu, 
and Jan Winter for helpful discussions.
Fermilab is operated by Fermi Research Alliance, LLC, under Contract
DE-AC02-07CH11359 with the United States Department of Energy.


\begin{thebibliography}{99}
%


\bibitem{Hubisz:2008gg}
  J.~Hubisz, J.~Lykken, M.~Pierini and M.~Spiropulu,
  {\it Phys.\ Rev.}\  {\bf D78} (2008) 075008
  [arXiv:0805.2398 [hep-ph]].


\bibitem{LHCinverse}
  P.~Binetruy, G.~L.~Kane, B.~D.~Nelson, L.~T.~Wang and T.~T.~Wang,
 {\it  Phys.\ Rev.}\   {\bf D70}  (2004) 095006
  [arXiv:hep-ph/0312248];
  J.~L.~Bourjaily, G.~L.~Kane, P.~Kumar and T.~T.~Wang,
  arXiv:hep-ph/0504170;
  N.~Arkani-Hamed, G.~L.~Kane, J.~Thaler and L.~T.~Wang,
  {\it JHEP} {\bf 0608}  (2006) 070
  [arXiv:hep-ph/0512190].



\bibitem{Morrissey:2009tf}
  D.~E.~Morrissey, T.~Plehn and T.~M.~P.~Tait,
  arXiv:0912.3259 [hep-ph].



\bibitem{Chung:2003fi}
  D.~J.~H.~Chung, L.~L.~Everett, G.~L.~Kane, S.~F.~King, J.~D.~Lykken and L.~T.~Wang,
  {\it Phys.\ Rep.}\  {\bf 407} (2005) 1 
  [arXiv:hep-ph/0312378].



\bibitem{Brignole:1993dj}
  A.~Brignole, L.~E.~Ibanez and C.~Munoz,
  {\it Nucl.\ Phys.}\   {\bf B422} (1994) 125 
  [Erratum-ibid.\   {\bf B436} (1995) 747 ]
  [arXiv:hep-ph/9308271].



\bibitem{Acharya:2008zi}
  B.~S.~Acharya, K.~Bobkov, G.~L.~Kane, J.~Shao and P.~Kumar,
  {\it Phys.\ Rev.}\   {\bf D78}  (2008) 065038
  [arXiv:0801.0478 [hep-ph]].



\bibitem{Randall:1998uk}
  L.~Randall and R.~Sundrum,
  {\it Nucl.\ Phys.}\   {\bf B557}  (1999) 79
  [arXiv:hep-th/9810155].


\bibitem{msugra}
  A.~H.~Chamseddine, R.~Arnowitt and P.~Nath,
  {\it Phys.\ Rev.\ Lett.}\  {\bf 49}  (1982) 970;
  R.~Barbieri, S.~Ferrara and C.~A.~Savoy,
 {\it  Phys.\ Lett.}\  {\bf B119}  (1982) 343;
  L.~J.~Hall, J.~D.~Lykken and S.~Weinberg,
  {\it Phys.\ Rev.}\  {\bf D27} (1983) 2359.


\bibitem{Ellis:2003dn}
  J.~R.~Ellis, K.~A.~Olive, Y.~Santoso and V.~C.~Spanos,
  {\it Phys.\ Lett.}\   {\bf B588} (2004) 7 
  [arXiv:hep-ph/0312262].


\bibitem{Cremmer:1983bf}
  E.~Cremmer, S.~Ferrara, C.~Kounnas and D.~V.~Nanopoulos,
  {\it Phys.\ Lett.}\   {\bf B133}  (1983) 61.
  J.~R.~Ellis, A.~B.~Lahanas, D.~V.~Nanopoulos and K.~Tamvakis,
  {\it Phys.\ Lett.}\   {\bf B134}  (1984) 429.



\bibitem{Giudice:1998bp}
  G.~F.~Giudice and R.~Rattazzi,
  {\it Phys.\ Rep.}\  {\bf 322} (1999) 419 
  [arXiv:hep-ph/9801271].



\bibitem{gaugoinomed}
  D.~E.~Kaplan, G.~D.~Kribs and M.~Schmaltz,
 {\it  Phys.\ Rev.}\   {\bf D62}  (2000) 035010
  [arXiv:hep-ph/9911293];
  Z.~Chacko, M.~A.~Luty, A.~E.~Nelson and E.~Ponton,
  {\it JHEP} {\bf 0001} (2000) 003 
  [arXiv:hep-ph/9911323].


\bibitem{Chacko:2000fn}
  Z.~Chacko and M.~A.~Luty,
  {\it JHEP} {\bf 0105}  (2001) 067
  [arXiv:hep-ph/0008103].



\bibitem{Kane:1993td}
  G.~L.~Kane, C.~F.~Kolda, L.~Roszkowski and J.~D.~Wells,
  {\it Phys.\ Rev.}\   {\bf D49}  (1994) 6173
  [arXiv:hep-ph/9312272].

\bibitem{Berger:2008cq}
  C.~F.~Berger, J.~S.~Gainer, J.~L.~Hewett and T.~G.~Rizzo,
  {\it JHEP} {\bf 0902} (2009) 023 
  [arXiv:0812.0980 [hep-ph]].




\bibitem{LH}
  M.~Schmaltz and D.~Tucker-Smith,
  {\it Annu.\ Rev.\ Nucl.\ Part.\ Sci.}\  {\bf 55}  (2005) 229
  [arXiv:hep-ph/0502182];
  M.~Perelstein,
  {\it Prog.\ Part.\ Nucl.\ Phys.}\  {\bf 58}  (2007) 247
  [arXiv:hep-ph/0512128].


\bibitem{Chacko:2005pe}
  Z.~Chacko, H.~S.~Goh and R.~Harnik,
  {\it Phys.\ Rev.\ Lett.}\  {\bf 96}  (2006) 231802
  [arXiv:hep-ph/0506256].




\bibitem{Hill:2002ap}
  C.~T.~Hill and E.~H.~Simmons,
  {\it Phys.\ Rep.}\  {\bf 381} (2003) 235 
  [Erratum-ibid.\  {\bf 390}  (2004) 553]
  [arXiv:hep-ph/0203079].



\bibitem{warped}
  L.~Randall and R.~Sundrum,
  {\it Phys.\ Rev.\ Lett.}\  {\bf 83}  (1999) 3370
  [arXiv:hep-ph/9905221].
  L.~Randall and R.~Sundrum,
  {\it Phys.\ Rev.\ Lett.}\  {\bf 83} (1999) 4690 
  [arXiv:hep-th/9906064].
  J.~D.~Lykken and L.~Randall,
  {\it JHEP} {\bf 0006}  (2000) 014
  [arXiv:hep-th/9908076].
  H.~Davoudiasl, J.~L.~Hewett and T.~G.~Rizzo,
  {\it Phys.\ Lett.}\   {\bf B473}  (2000) 43
  [arXiv:hep-ph/9911262].


\bibitem{Csaki:2005vy}
  C.~Csaki, J.~Hubisz and P.~Meade,
  arXiv:hep-ph/0510275.


\bibitem{ArkaniHamed:1998rs}
  N.~Arkani-Hamed, S.~Dimopoulos and G.~R.~Dvali,
  {\it Phys.\ Lett.}\   {\bf B429}  (1998) 263
  [arXiv:hep-ph/9803315].


\bibitem{UED}
T.~Appelquist, H.~C.~Cheng and B.~A.~Dobrescu,
{\it Phys.\ Rev.}\   {\bf D64}  (2001) 035002;
  H.~C.~Cheng, K.~T.~Matchev and M.~Schmaltz,
  {\it Phys.\ Rev.}\   {\bf D66}  (2002) 036005
  [arXiv:hep-ph/0204342];
G.~Servant and T.~M.~P.~Tait,
 {\it Nucl.\ Phys.}\   {\bf B650}  (2003) 391
 [arXiv:hep-ph/0206071];
 H.~C.~Cheng, J.~L.~Feng and K.~T.~Matchev,
{\it Phys.\ Rev.\ Lett.}\  {\bf 89}  (2002) 211301
[arXiv:hep-ph/0207125].


\bibitem{Strassler:2006im}
  M.~J.~Strassler and K.~M.~Zurek,
  {\it Phys.\ Lett.}\   {\bf B651}  (2007) 374
  [arXiv:hep-ph/0604261].


\bibitem{Georgi:2007ek}
  H.~Georgi,
  {\it Phys.\ Rev.\ Lett.}\  {\bf 98}  (2007) 221601
  [arXiv:hep-ph/0703260].


\bibitem{Bai:2009ms}
  Y.~Bai, M.~Carena and J.~Lykken,
  {\it Phys.\ Rev.\ Lett.}\  {\bf 103}  (2009) 261803
  [arXiv:0909.1319 [hep-ph]].


  \bibitem{ArkaniHamed:2008qn}
 N.~Arkani-Hamed, D.~P.~Finkbeiner, T.~R.~Slatyer and N.~Weiner,
  {\it Phys.\ Rev.}\   {\bf D79}  (2009) 015014
  [arXiv:0810.0713 [hep-ph]].


\bibitem{pdg}
C. Amsler {\it et al.} (Particle Data Group), Phys. Lett. {\bf B667} (2008) 1.



 \bibitem{DeRujula:2010ys}
  A.~De Rujula, J.~Lykken, M.~Pierini, C.~Rogan and M.~Spiropulu,
  arXiv:1001.5300 [hep-ph].

\bibitem{Gao:2010qx}
  Y.~Gao, A.~V.~Gritsan, Z.~Guo, K.~Melnikov, M.~Schulze and N.~V.~Tran,
  {\it Phys.\ Rev.}\   {\bf D81}  (2010) 075022
  [arXiv:1001.3396 [hep-ph]].



\bibitem{Cheng:2002ab}
  H.~C.~Cheng, K.~T.~Matchev and M.~Schmaltz,
  {\it Phys.\ Rev.}\   {\bf D66}  (2002) 056006
  [arXiv:hep-ph/0205314].

\bibitem{Datta:2005zs}
  A.~Datta, K.~Kong and K.~T.~Matchev,
  {\it Phys.\ Rev.}\   {\bf D72}  (2005) 096006
  [Erratum-ibid.\   {\bf D72}  (2005) 119901]
  [arXiv:hep-ph/0509246].


\bibitem{Hallenbeck:2008hf}
  G.~Hallenbeck, M.~Perelstein, C.~Spethmann, J.~Thom and J.~Vaughan,
  {\it Phys.\ Rev.}\   {\bf D79}  (2009) 075024
  [arXiv:0812.3135 [hep-ph]].


\bibitem{ArkaniHamed:2001nc}
  N.~Arkani-Hamed, A.~G.~Cohen and H.~Georgi,
  {\it Phys.\ Lett.}\   {\bf B513}  (2001) 232
  [arXiv:hep-ph/0105239].




\end{thebibliography}
\end{document}